\documentclass[conference]{IEEEtran}
\IEEEoverridecommandlockouts

\usepackage[normalem]{ulem} 
\usepackage{xcolor} 
\usepackage{graphicx} 
\usepackage{epstopdf} 
\usepackage{subfigure}

\usepackage{xcolor} 

\usepackage{cite}
\usepackage{amsmath,amssymb,amsfonts,dsfont}
\usepackage{algorithmic}
\usepackage{graphicx}
\usepackage{textcomp}
\usepackage{xcolor}
\newtheorem{definition}{Definition}
\newtheorem{assumption}{Assumption}
\newtheorem{lemma}{Lemma}
\newtheorem{theorem}{Theorem}

\newtheorem{corollary}{Corollary}
\newtheorem{proposition}{Proposition}

\newtheorem{remark}{Remark}
\newcommand*{\qed}{\null\nobreak\hfill\ensuremath{\square}}%

\def\BibTeX{{\rm B\kern-.05em{\sc i\kern-.025em b}\kern-.08em
    T\kern-.1667em\lower.7ex\hbox{E}\kern-.125emX}}
\begin{document}

\title{\vspace{0.5cm}Deception in Asymmetric Information Homicidal Chauffeur Game\\
}

\author{Shreesh Mahapatra$^1$
\and 
 Bhargav Jha$^2$

\and
Michael R.~Dorothy$^3$

\and 
 Shaunak D.~Bopardikar$^4$

 \thanks{This work was sponsored by the Army Research Laboratory and was accomplished under Cooperative Agreement Number W911NF-17-2-0181. The views expressed in this paper are those of the authors and do not reflect the official policy or position of the United States Government, Department of Defense, or its components. This work was also partially supported by the Anusandhan National Research Foundation's Prime Minister Early Career Grant number ANRF/ECRG/2024/005359/ENS}
\thanks{$^{1}$Shreesh Mahapatra is with the Department of Mechanical Engineering, Indian Institute of Technology, Kharagpur, WB, 721302, India. Email: {\tt \small shreesh@kgpian.iitkgp.ac.in}}%
\thanks{$^{2}$Bhargav Jha is with the Department of Electrical Engineering, Indian Institute of Technology, Kharagpur, WB, 721302, India. Email: {\tt \small bhargav@ee.iitkgp.ac.in}}%
\thanks{$^{3}$Michael Dorothy is with the Army Research Directorate, DEVCOM Army Research Laboratory, APG, MD, 20783, USA. Email: {\tt \small michael.r.dorothy.civ@army.mil} }%
\thanks{$^{4}$Shaunak D. Bopardikar is with the Department of Electrical and Computer Engineering, Michigan State University. Email: {\tt\small shaunak@egr.msu.edu}}
}

\maketitle

\begin{abstract}
The classic Homicidal Chauffeur game is a pursuit-evasion game played in an unbounded planar environment between a pursuer constrained to move with fixed speed on curves with bounded curvature, and a slower evader with fixed speed but with simple kinematics. We introduce a new variant of this game with asymmetric information in which the evader has the ability to choose its speed among a finite set of choices that is unknown to the pursuer a priori. Therefore the pursuer is required to estimate the evader's maximum speed based on the observations so far. This formulation leads to the question of whether the evader can exploit this asymmetry by moving deceptively by first picking a slower speed to move with and then switching to a faster speed when a specified relative configuration is attained to increase the capture time as compared to moving with the maximum speed at all times. Our contributions are as follows. First, we derive optimal feedback Nash equilibrium strategies for the complete information case of this game in which the evader is allowed to vary its speed in a given interval. Second, for the version with asymmetric information, we characterize regions of initial player locations in the game space from which the evader does not have any advantage in using deceptive strategies. Finally, we provide numerical evidence of regions in the game space from which the evader can increase the capture time by moving deceptively.
\end{abstract}

\begin{IEEEkeywords}
Game-theory, Deception, Pursuit-evasion games
\end{IEEEkeywords}

\section{Introduction}
The Homicidal Chauffeur game involves a pursuer that is constrained to move with fixed speed on curves with bounded curvature and an evader that is slower than the pursuer, but does not have any turning constraints\cite{isaacs1999differential}. The evader is captured once the distance to the pursuer falls below a threshold. Although classical solutions exist \cite{isaacs1999differential,merz1974homicidal,pachter2019classical}, this model has gained renewed attention for applications in autonomous vehicle planning \cite{exarchos2015suicidal} and search–tracking missions with ground or aerial vehicles \cite{chaudhari2021time}. This paper explores a new facet of the Homicidal Chauffeur game: the role of information on the evader’s maximum speed and whether moving slower, thereby concealing true capability, can prolong capture.

\medskip

It is well known that players’ information sets strongly influence Nash equilibria in non-cooperative games \cite{bacsar1998dynamic}, and partial information adds complexity due to \emph{dueling estimators}. Still, deception has been modeled in security contexts \cite{fuchs2011}, through asymmetric sensing \cite{brooks2008game}, asset revelation \cite{hespanha2000deception}, or noisy sensors \cite{hespanha2019sensor}. Other works study repeated zero-sum games with asymmetric knowledge \cite{jones2012policy,aziz2019smart}, while bounded rationality models \cite{fotiadis2023game} capture how agents estimate opponents’ reasoning levels.

\medskip

Deception in pursuit–evasion games has largely been studied in specific contexts. Early work examined errors in pursuer bearing measurements \cite{yavin1987pursuit}, while others investigated motion camouflage strategies that exploit limitations in evader perception \cite{justh2006steering,reddy2007motion}. More recently, \cite{shishika2024deception} studied a differential game where attackers deceive a turret to improve their chances of reaching a perimeter. Beyond these, to our knowledge, no prior work has considered \emph{deception through motion}.

\medskip

In this paper, we introduce a new variant of the Homicidal Chauffeur game where the evader can choose between two speeds, while the pursuer does not know the evader’s maximum speed and must estimate it from observations. We ask whether the evader can exploit this asymmetry by initially moving slowly and later switching to a faster speed to prolong capture. Our contributions are as follows. First, derivation of optimal feedback Nash equilibrium strategies for the complete information case with variable evader speed. Second, characterization of regions where deception offers no advantage. Finally, numerical evidence of regions where deception successfully increases capture time.

\section{Preliminaries: Full Information Homicidal Chauffeur Game}
Before we pose the asymmetric information homicidal chauffeur game (HCG), we will present a summary of the full information version of the HCG as a preliminary. 
\subsection{Problem Formulation}
\begin{figure}[!htbp]
    \centering
    \includegraphics[width=0.4\textwidth]{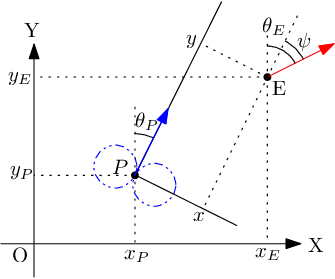}
    \caption{Engagement Kinematics}
    \label{engagmentKinematics}
\end{figure}

Consider a planar engagement in a Cartesian coordinate system ($X$-$O$-$Y$) with a pursuer $(P)$ and an evader $(E)$ as shown in Fig. \ref{engagmentKinematics}.  The pursuer is a constant speed, Dubins vehicle whose states are its planar position $\left(x_P(t),~y_P(t)\right)\in\mathds{R}^2$ and heading angle $\theta_P(t) \in [-\pi,\pi]$, and the control input is the bounded turn-rate $\left|u(t)\right|\leq 1$. In the context of the paper, it can be assumed, without loss of generality that the turn radius and the speed of the pursuer are both normalized to unity. The evader has simple motion kinematics with its planar position as states $\left(x_E(t),~y_E(t)\right)\in\mathds{R}^2$ and its heading angle $\theta_E(t)\in[-\pi,\pi]$ as the control input. Unless stated otherwise, we follow the same convention as Isaacs \cite{isaacs1999differential} and measure all angles from the positive $Y$-axis in the clockwise direction. The kinematics of both the pursuer and the evader are governed by the following set of differential equations,
\begin{subequations}\label{fullOrderKinematics}
\begin{align}
    &\dot{x}_E(t) = \mu \sin\theta_E(t), ~~\dot{y}_E(t) = \mu \cos\theta_E(t), \nonumber \\ &~~~~~~~~~~~~~~x_E(0) = x_{E0},~ y_E(0)=y_{E0}, \\
    &\dot{x}_P(t) = \sin\theta_P(t), ~~\dot{y}_P(t) = \cos\theta_P(t),~~ \dot{\theta}_P(t) = u(t), \nonumber \\ &~~~~~~~~~~~~~~x_P(0)=x_{P0},~ y_P(0)=y_{E0},~ \theta_P(0) = \theta_{P0},
\end{align}
\end{subequations}
where $\left(x_{P0},y_{P0}, \theta_{P0}\right)$ and $(x_{E0},y_{E0})$ are the initial conditions for $P$ and $E$, respectively. Let the terminal time of capture be defined as the time instance $t_f$ at which 
\[ 
\left(x_P(t_f)-x_E(t_f)\right)^2+\left(y_P(t_f)-y_E(t_f)\right)^2 \leq l^2, \]
where $l$ is the capture radius. For the sake of brevity, unless stated otherwise, we will drop the notation for the time dependency of the states and control inputs. 

\medskip

The following are the assumptions in the full information HCG.

\begin{assumption}\label{as:speed}
The pursuer has a strict speed advantage over the evader, \textit{i.e.} the speed of the evader, $\mu\in[0,1)$.
\end{assumption}

\begin{assumption}\label{as:fullinfo}
Both $P$ and $E$ have complete information of the states $(x_P,y_P,\theta_P,x_E,y_E)$ and the parameters $\left(\mu , l \right)$.
\end{assumption}

\begin{assumption}\label{as:classical}
The parameters correspond to the classical case of the HCG as discussed in \cite{isaacs1999differential,pachter2019classical}. Hence, $\mu^2 + l^2<1$.
\end{assumption}

\medskip

$P$ and $E$ play a zero-sum minimax differential game where $P$ minimizes the time to capture $t_f$, while $E$ maximizes it. Hence, the objective of the game is defined as:
\begin{equation}
    \underset{u\in[-1,1]}{\min}~ \underset{\theta_E\in[-\pi,\pi]}{\max}~ J \triangleq t_f
\end{equation}

 \begin{lemma} \cite{isaacs1999differential}
 Under Assumptions~\ref{as:speed}, \ref{as:fullinfo} and \ref{as:classical}, the time of capture is finite, \textit{i.e.} $t_f < \infty$. 
 \end{lemma}
 
\subsection{Pursuer Relative Coordinates}
To reduce the number of states, we consider a coordinate system with $P$ at the origin and the $Y$-axis being always aligned with the pursuer's heading $\theta_P$. The position of $E$ in this coordinate system is given by $(x,y)\in\mathds{R}^2$ and the relative heading of $E$ with respect to $P$'s heading is $\psi:=\theta_E-\theta_P$. The reduced order kinematics with states $\mathbf{x}=[x~y]^\top$ and control inputs $u$ and $\psi$ for $P$ and $E$, respectively, is given by, 
\begin{align}\dot{\mathbf{x}} = f(\mathbf{x},u,\psi) \label{dynamicsRel} \end{align} where 
\begin{align*}
    f(\mathbf{x},u,\psi) = \begin{bmatrix} -y u + \mu \sin \psi \\ 
      x u -1 + \mu \cos \psi \end{bmatrix}, &|u|\leq 1,~ \psi \in [-\pi,\pi] 
\end{align*}

Note that reducing the number of states to two is possible as $E$ can command its heading instantaneously. Readers interested in details should refer to \cite{bacsar1998dynamic,isaacs1999differential}. The game can now be expressed as 
\begin{equation}
    \underset{u\in[-1,1]}{\min}~ \underset{\psi\in[-\pi,\pi]}{\max}~ J \triangleq t_f.
\end{equation}
Further, the equivalent boundary of the terminal capture set $(\mathcal{C})$ in the reduced coordinates is defined as $\mathcal{C}= (x,y):x^2+y^2 = l^2$. It is known from \cite{isaacs1999differential}, that the capture occurs at this boundary.

The solution of the HCG has a mirror symmetry about the $Y$-axis. Hence, for brevity, we will discuss the results only for the case when $x\geq 0$.  
 \begin{figure}[htbp]
     \centering 
     \hspace*{-3.2cm}
\includegraphics[width=0.8\textwidth]{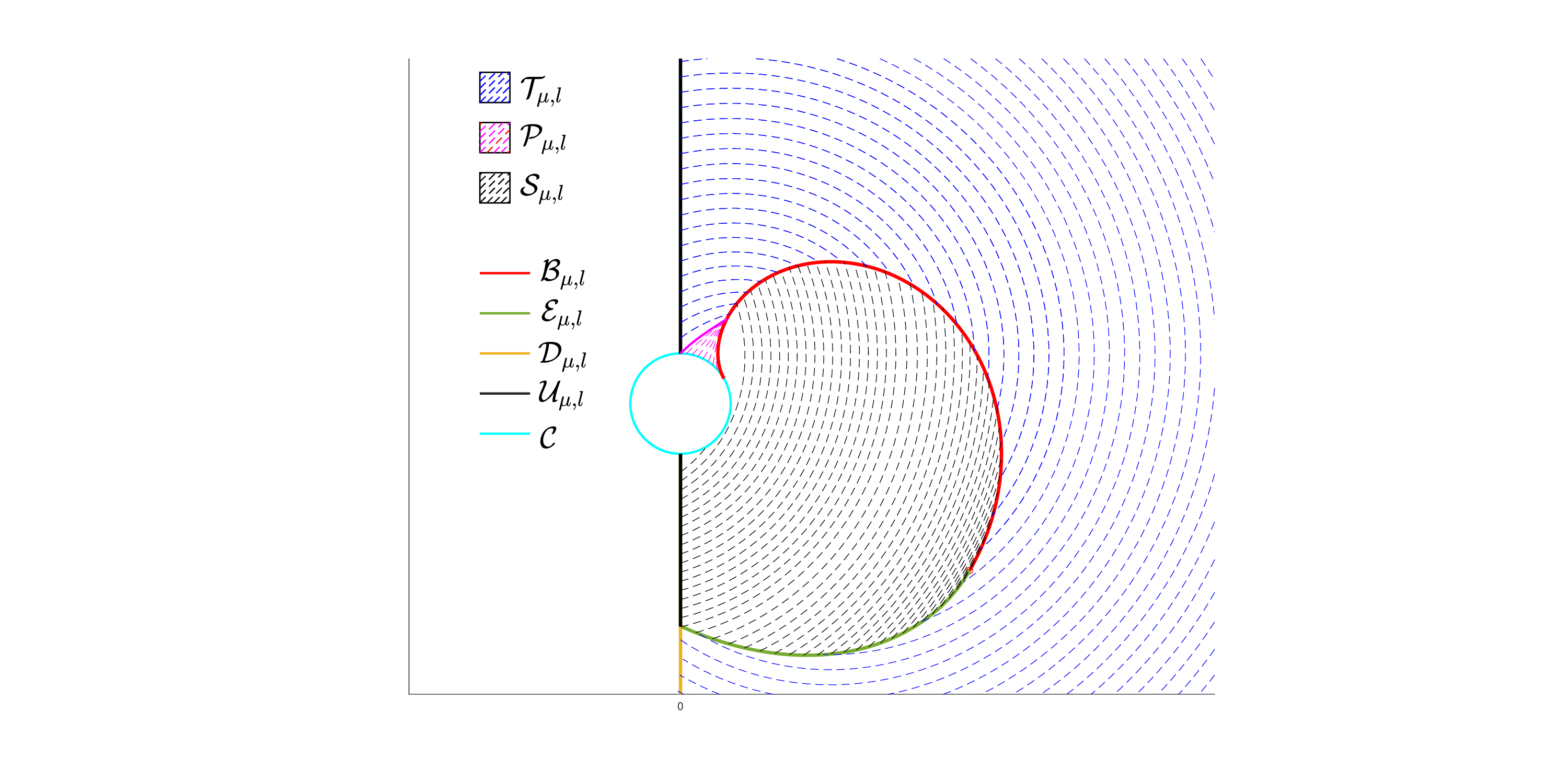}
\caption{Regions in the game-space of HCG for $x>0$}
\end{figure}
\begin{definition}[Feedback Equilibrium Strategy] A pair of feedback strategies \( \left(u^*(x,y),\psi^*(x,y)\right) \) is a \emph{feedback equilibrium strategy} if and only if
\begin{multline}
J\left(u^*(x,y),\psi(x,y)\right)\leq J\left(u^*(x,y),\psi^*(x,y)\right) \\ \leq J\left(u(x,y),\psi^*(x,y)\right),
\end{multline}
for any $\psi(x,y)\in[-\pi,\pi]$ and $u(x,y)\in[-1,1]$.
 \end{definition} 

 \medskip
 
\begin{definition}[Equilibrium trajectory] A trajectory $\left(x(t),y(t)\right),~~\forall t\in[0,t_f]$, generated by a pair of equilibrium strategies is defined as an \emph{equilibrium trajectory}. 
 \end{definition}

 \medskip
 
 \begin{definition}[Usable part (UP)] For a given $(\mu,l)$, the \emph{usable part} is the set of points $(x,y)\in \mathcal{C}$, where the equilibrium trajectories terminate.  In the context of the full information HCG, the usable part is the part of the capture set such that $\phi \in  \left(-\cos^{-1}\mu,\cos^{-1}\mu\right)$. The boundary points on the UP corresponding to $\phi = \pm \cos^{-1}\mu$ are called boundary of usable part (BUP). 
\end{definition}

\medskip

\begin{definition}[Primary trajectory region $\left(\mathcal{P}_{\mu,l}\right)$] For a given $(\mu,l)$, the \emph{primary trajectory region} is the set of all initial positions from which there exists an equilibrium trajectory that terminates on the usable part such that $\phi\in\left(0,\cos^{-1}\mu\right)$.  An  equilibrium trajectory in this region is called the \emph{primary equilibrium trajectory} and the corresponding feedback equilibrium strategy is given by
\begin{align}
    \left(u^*, \psi^*\right) = \left(1, \phi + \tau\right),  \phi\in\left(0,\cos^{-1}\mu\right)
\end{align}
where $\tau:=t_f-t$ denotes the time to capture.
\end{definition}

\medskip

\begin{definition}[Barrier curve $\left(\mathcal{B}_{\mu,l}\right)$] \label{def:barrier} The \emph{barrier curve} is a set of points that separates the game space into two regions, one in which $u^*=1$ and the other in which $u^*=-1$. The barrier curve is tangent to the capture set at the point $l\sin(\cos^{-1}\mu),l\cos(\cos^{-1}\mu)$.
\end{definition}

\medskip

\begin{definition}[Positive universal region $\left(\mathcal{U}^{+}_{\mu,l}\right)$] For a given $(\mu,l)$, the \emph{positive universal region} is the set of points $(x,y): x=0, y>l$. On the positive universal region, $u^*\equiv 0$. Here, $E$ is in front of $P$.
\end{definition}

\medskip

\begin{definition}[Tributary trajectory region $\left(\mathcal{T}_{\mu,l}\right)$]\label{deftot} For a given $(\mu,l)$, the \emph{tributary trajectory region} is the set of all initial positions from which there exists an equilibrium trajectory  with the equilibrium strategy
   \begin{align}
    \left(u^*, \psi^*\right) = \left(1, \tau\right), \quad \tau \geq 0.
\end{align}
A trajectory with the above equilibrium strategy is called a \emph{tributary trajectory}. A tributary trajectory merges with the positive universal line.
\end{definition}

\medskip

\begin{definition}[Equivocal Curve ($\mathcal{E}_{\mu,l}$)] The \emph{equivocal curve} is a curve in the game-space where $E$ can choose from two distinct optimal strategies at each point. It can either traverse along the curve heading to end point of $\mathcal{B}_{\mu,l}$ or depart along $\mathcal{T}_{\mu,l}$. Along this curve, $P$'s optimal control may not be at the extremal and is dependent on the strategy of $E$. The strategy of $E$ is to pursuer $P$ using pure-pursuit geometrical rule.
\end{definition}

\medskip

\begin{definition}[Negative universal region  $\left(\mathcal{U}^{-}_{\mu,l}\right)$] For a given $(\mu,l)$, the \emph{negative universal region} is the set of points $(x,y): x=0, y\in(\bar{y}_{ES},-l)$, where $(0,\bar{y}_{ES})$ denotes the point where the equivocal curve meets the $y$-axis. On the negative universal region, $u^*\equiv 0$. Here, $P$ is in front of $E$.
\end{definition}

\medskip

\begin{definition}[Secondary trajectory region ($\mathcal{S}_{\mu,l}$)] For a given $(\mu,l)$, the \emph{secondary trajectory region} is the set of all initial positions from which there exists an equilibrium trajectory that can merge with $\mathcal{E}_{\mu,l}$ or $\mathcal{U}^-_{\mu,l}$. {If the equilibrium strategy merges with $\mathcal{E}_{\mu,l}$, then that equilibrium strategy is given by} 
\begin{align}
    \left(u^*, \psi^*\right) = \left(-1, \pi-\tau-\tan^{-1}(y_{ES}/x_{ES})\right), \quad \tau \geq 0,
\end{align}
where $(x_{ES}, y_{ES})$ is the point at which it intersects with $\mathcal{E}_{\mu,l}$. {If the equilibrium strategy merges with $\mathcal{U}^-_{\mu,l}$, then that equilibrium strategy is given by} 
\begin{align}
    \left(u^*, \psi^*\right) = \left(-1, -\tau\right), \quad \tau \geq 0.
\end{align}
\end{definition}

\medskip

\begin{definition}[Dispersal line ($\mathcal{D}_{\mu,l}$)] The \emph{dispersal line} is a set of points $\{(x,y): x=0, y\leq y_{ES}\}$
, where \( y_{ES} \) denotes the y-coordinate of the ES when it contacts the y-axis. Here, the players' equilibrium strategies are not unique.
\end{definition}

\medskip

\section{Geometric Analysis of HCG With the Variation in Evader's Speed Parameter}

Now we consider two different HCGs that correspond to the parameters $(\mu_1,l)$ and $(\mu_2,l)$, where $\mu_1>\mu_2$. We now propose some properties of the  games that arise due to the variation in the evader's speed.

\medskip
\subsection{Monotonicity of Capture Time with Evader's Speed}
The ensuing theorem characterizes the region in the game space in which the evader has no incentive to move at any slower speed.
\begin{theorem}\label{theorem1}
For any $(x_0,y_0)\in\mathcal{T}_{\mu_1,l}\cap \mathcal{T}_{\mu_2,l}$, let $t^i_f$ denote the time of capture for the HCG corresponding to the speed $\mu_i$. If $\mu_1>\mu_2$, then  $t^1_f > t^2_f$.  
\end{theorem}

\medskip


{The proof of Theorem \ref{theorem1} requires the following intermediate results.}

\begin{lemma} \label{line}
Consider a point $(x_0,y_0) \in \mathcal{T}_{\mu,l}$. For $E$, the equilibrium trajectory is a straight line that passes through the corresponding real coordinates of $E$ at $(x_{E0},y_{E0})$
\end{lemma}

\emph{Proof of Lemma~\ref{line}}: From the solution of the HCG and Definition \ref{deftot}, the equilibrium strategy for $P$ is $u^*=1$  and $E$ is $\psi^*=\tau$. Therefore, $P$ goes on a circular trajectory and its heading is given by 
    $\theta_P = \theta_{P0} + t$. In the global coordinate system, the heading angle for $E$ is expressed as $\theta_E = \theta_P + \psi^*$. As $\tau = t_f-t$, we have 
\begin{align*}
    \theta_E = \theta_P + \tau = \theta_{P0} + t + t_f - t = \theta_{P0} + t_f.
\end{align*}

\begin{figure}[!htbp]
     \centering        
        \includegraphics[width=0.4\textwidth]{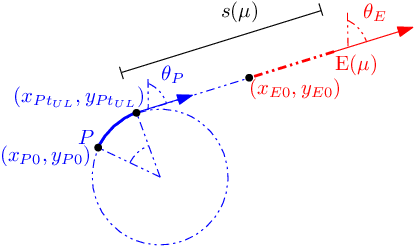}
        \caption{The CS trajectory for $P$ in $\mathcal{T}_{\mu,l}$}
        \label{cs}
\end{figure}

Thus $E$ maintains a constant heading and moves along a straight-line path starting from $(x_{E0},y_{E0})$, as illustrated in Fig. \ref{cs}.
This completes the proof of Lemma~\ref{line}.  \qed 

\medskip

\begin{lemma} \label{csdub}
For a given $(\mu,l)$, consider a point $(x_0,y_0)$ in the region $\mathcal{T}_{\mu,l}$. Let $\Theta(t)=(x_P(t),y_P(t)) \quad t\in[0,\bar{t}$] be the time-optimal trajectory for $P$ to traverse from the configuration $\left(x_{P0},y_{P0},\theta_{P0}\right)$ to the point $(x_{E0},y_{E0})$ \cite{dubins1957curves,boissonnat1994accessibility}. The trajectory $\Theta(t)$ is a subset of the equilibrium trajectory for $P$. 
\end{lemma}

\medskip

\emph{Proof of Lemma~\ref{csdub}:} From Lemma \ref{line} and the solution of the HCG in $\mathcal{T}_{\mu,l}$, we observe that $P$ must first follow a circular trajectory of turn radius $1$ until the time it reaches the $\mathcal{U}^+_{\mu,l}$ (say $(t_{UL})$) arriving at the point $(x_{Pt_{UL}}, y_{Pt_{UL}})$ as depicted in Fig. \ref{cs}. Now as $\psi=0$ on the $\mathcal{U}^+_{\mu,l}$, $P$ pursues $E$ along a straight line path with $ \theta_E(t) = \theta_P(t) \quad \forall t\in[t_{UL}, t_f]$. As illustrated in Fig. \ref{cs}, this implies that the straight line path passes through $(x_{E0},y_{E0})$. This completes the proof of Lemma~\ref{csdub}. \qed

\medskip

\begin{corollary}\label{unis}
Consider ${\mu}_1 > {\mu}_2$ and a point $(x_0,y_0)\in\mathcal{T}_{\mu_1,l}\cap\mathcal{T}_{\mu_2,l}$. If $t^i_{UL}$ denotes the time-to-reach $\mathcal{U}^+_{\mu_i,l}$ from $(x_0,y_0)$ along the equilibrium trajectory of the HCG with parameters $(\mu_i,l)$, then $t^1_{UL}=t^2_{UL}$.  
\end{corollary}

\emph{Proof.} From Lemma~\ref{csdub}, $P$ moves on a circular path until aligned with $(x_0,y_0)$ (Fig.~\ref{cs}). Thus, the time to reach $\mathcal{U}^+_{\mu,l}$ is independent of $\mu_i$, completing the proof. \qed

From Corollary~\ref{unis}, the separation after reaching the universal line is -
\begin{align*}
s(\mu) = s_0+\mu \, t_{UL},  
s_0=\sqrt{(x_{E0}-x_{Pt_{UL}})^2+(y_{E0}-y_{Pt_{UL}})^2}.
\end{align*}
Since $\mu_1>\mu_2$, we have $s(\mu_1)>s(\mu_2)$. The capture time after reaching $\mathcal{U}^+_{\mu,l}$ is
\[
\tau=\frac{s}{v_p(1-\mu)}.
\]
Because $s(\mu_1)>s(\mu_2)$ and $1-\mu_1<1-\mu_2$, it follows $\tau_1>\tau_2$, implying $t_F^1>t_F^2$. Hence, a faster $E$ prolongs capture, completing the proof of Theorem~1. \qed
\medskip

\subsection{Barrier Separation Property for Different Evader Speeds}

The next result establishes a property of the barrier curves from Definition~\ref{def:barrier}.

\medskip

\begin{theorem}\label{thm:barrier}
Let $\mathcal{B}(\mu_1,l)$ and $\mathcal{B}(\mu_2,l)$ be the barrier curves of the HCG corresponding to the evader speed $\mu_1$ and $\mu_2$, respectively. Then, 
\[\mathcal{B}(\mu_1,l)\cap \mathcal{B}(\mu_2,l)=\emptyset.\]
\end{theorem}

Although analytic expressions for the barrier were obtained in \cite{isaacs1999differential,pachter2019classical}, the non-intersection of the barriers for two different parameters can only be verified numerically by finding the roots of a nonlinear equation. In the rest of this section, we present the following analytic proof of Theorem~\ref{thm:barrier}.

\medskip

Consider a variant of HCG in which the evader's speed $\mu$ is an additional bounded control input, i.e., $\mu(t)\in(0,\bar{\mu}]$. As before, we assume that $P$ and $E$ have access to current states $\left(x(t),y(t)\right)$ and the parameters $(\bar{\mu},l$ of the game.  We term this game as the \emph{Homicidal Chauffeur Game with Varying Speed Evader (HCGVSE)}. The kinematics as well as the reduced-order kinematics of the HCGVSE are identical to those in Eqs. \eqref{fullOrderKinematics} and Eqs. \eqref{dynamicsRel}, respectively. The minimax game is now expressed as 
\begin{equation}
    \underset{u\in[-1,1]}{\min}~ \underset{\mu\in (0,\bar{\mu}], ~ \psi\in[-\pi,\pi]}{\max}~ \bar{J} \triangleq t_f.
\end{equation}

\medskip

{The following three results are intermediate properties that will be required to establish Theorem~\ref{thm:barrier}.}

\medskip

\begin{lemma}\label{upLemma}
In the HCGVSE, the capture will occur on the segment of $\mathcal{C}$ such that $$\left(x(t_f),y(t_f)\right) = (l \sin\phi, ~l \cos\phi), ~~~\forall \phi \in [0, \cos^{-1} \bar{\mu}].$$
\end{lemma}

\emph{Proof of Lemma~\ref{upLemma}:} Any $(x,y)\in \mathcal{C}$ can be parameterized as $(l \sin\phi, l \cos\phi), \quad \forall \phi\in [-\pi,\pi]$. The unit vector along the inward pointing normal  to the terminal set is given by $$\mathbf{\hat{n}} = [-\sin\phi,~ - \cos \phi].$$
At an infinitesimally small time before capture, the pursuer will maximize the flow of the game along $\mathbf{\hat{n}}$ and the evader will minimize it. Hence, we have 
\begin{align}
    &\underset{u\in[-1,1]}{\max}~ \underset{\mu\in [0,\bar\mu],\psi\in [-\pi,\pi]}{\min} \sin\phi (y(t_f) u - \mu \sin \psi) \nonumber  \\ & ~~~~~~~~~~~~~~~~- \cos\phi (x(t_f) u - 1 + \mu \cos\psi) \\
 &=\underset{\mu\in [0,\bar\mu],\psi\in [-\pi,\pi]}{\min} \cos\phi - \mu\cos(\phi-\psi),
\end{align}
where the parameterization $(x(t_f),y(t_f))=(l\sin\phi,l\cos\phi)$ is used.
As the capture is guaranteed, at the boundary of the usable portion of the capture set,
\begin{align}
     &\underset{\mu\in [0,\bar\mu],\psi\in [-\pi,\pi]}{\min} \cos\phi - \mu\cos(\phi-\psi) = 0
\end{align}
Therefore, $\phi\leq \cos^{-1}\bar{\mu}$, and that completes the proof of Lemma~\ref{upLemma}. \qed

\medskip

{The next result relates the barrier curves in the HCGVSE and the classic HCG}.

\medskip

\begin{lemma}\label{lem:barrier}
The barrier curve for the HCGVSE with parameters $(\bar{\mu},l)$ is identical to the barrier curve of the HCG with parameters $(\bar{\mu},l)$.
\end{lemma}

\medskip

\emph{Proof of Lemma~\ref{lem:barrier}:} Using the kinematics in Eq. \eqref{dynamicsRel}, the Hamiltonian can be written as
\begin{align}
    H = -1 + p_x(t)(- y u + \mu \sin \psi) + p_y(t)(x u -1 + \mu \cos \psi),
\end{align}
where $p_x(t)$ and $p_y(t)$ are the costates.
From the necessary conditions of the equilibrium strategy \cite[Theorem 2, Chapter 8]{bacsar1998dynamic} , we have that
as $P$ maximizes the Hamiltonian, it's strategy is given as,
\begin{align}
    u^* &= \underset{u\in[-1 ,1]}{\arg\max}~ H(\mathbf{x},u,\mu,\psi) = \mathrm{sign}(p_y(t)x - p_x(t) y). \end{align}
As $E$ minimizes the Hamiltonian, the equilibrium strategy of $E$ is given by 
\begin{align}
    (\mu^*,~\psi^*) = \underset{\mu\in[-1,1],~\psi\in [-\pi,\pi]}{\arg\min} H(\mathbf{x},u,\mu,\psi) 
    \end{align}
Hence, we have
    $\mu^* = \bar{\mu}$,  and $\psi^*$ such that 
    $$\sin(\psi^*) = -\frac{p_x(t)}{\sqrt{p^2_x(t)+p^2_y(t)}}, ~~ \cos(\psi^*) = -\frac{p_y(t)}{\sqrt{p^2_x(t)+p^2_y(t)}}$$
Using the costate equations and the transversality conditions, we have
$$p_x(t) = -\alpha \sin(\phi + t_f - t), \quad p_y(t) = -\alpha \cos(\phi + t_f - t) $$
where $\alpha>0$ and $\phi\in[0,\cos^{-1}\mu)$. Hence, we have 
$$\sin(\psi^*) = \sin(\phi+\tau), ~~ \cos(\psi^*) = \cos(\phi+\tau)$$ where $\tau = t_f-t$.

From Lemma \ref{upLemma}, the BUP of HCG is identical to the HCG with varying speed. Substituting the above equilibrium control inputs and integrating dynamical equations \eqref{dynamicsRel} with initial conditions starting at the BUP, we obtain that the barrier in HCG with varying speed is identical to  version of HCG with constant speed. This completes the proof of Lemma~\ref{lem:barrier}. \qed

\medskip

{To state the next intermediate result, we recall the concept of a \emph{semipermeable surface} in dynamic games \cite{isaacs1999differential}.}

\medskip

\begin{definition}[Semipermeable Surface] Let $\mathcal{D}(\mathbf{x}) = 0$ be a surface in the game space and $\mathbf{n}_{\mathcal{D}}(\mathbf{x})$ be the normal to $\mathcal{D}(\mathbf{x})$ at $\mathbf{x}$. The surface is termed as \emph{semipermeable} if at each point $\mathbf{x}\in\mathcal{D}(\mathbf{x})$, there exist a $(u^*,\mu^*,\psi^*)$ such that
\begin{align*}
&f(\mathbf{x},u^*,\mu,\psi).\mathbf{n}_{\mathcal{D}}(\mathbf{x}) \leq f(\mathbf{x},u^*,\mu^*,\psi^*).\mathbf{n}_{\mathcal{D}}(\mathbf{x}) = 0 \\ &f(\mathbf{x},u^*,\mu^*,\psi^*).\mathbf{n}_{\mathcal{D}}(\mathbf{x}) = 0 \leq  f(\mathbf{x},u,\mu^*,\psi^*).\mathbf{n}_{\mathcal{D}}(\mathbf{x}), \nonumber \\ & \quad \quad \quad \quad \quad \forall u\in[-1,1], \psi\in[0,\bar{\mu}], \psi\in[-\pi,\pi].
\end{align*}
\end{definition}
In other words, a semipermeable surface partitions the game space into two regions, namely a $P-$region and a $E-$region. By definition, at any point in the $P$(resp., $E$) region, if the pursuer (resp., evader) applies its optimal strategy, then there does not exist a strategy for $E$(resp., $P$) that drives the trajectory from the $P-$region (resp., $E$-region) to the $E-$region (resp., $P-$region).

\begin{lemma}\label{semipermeableBarrier}
The barrier of the HCGVSE is a semipermeable surface.  
\end{lemma}

\emph{Proof of Lemma~\ref{semipermeableBarrier}:} The proof directly follows from {Lemma~\ref{lem:barrier} which established that the barrier surfaces in the HCG and the HCGVSE are identical,} and the steps outlined in \cite[Section 8.2.2]{lewin2012differential} {which show that the barrier surface of the HCG is semipermeable.} \qed

\medskip

We are now ready to present the proof of Theorem~\ref{thm:barrier}.

\medskip

\begin{figure}[!htbp]
    \centering
    \includegraphics[width=0.35\textwidth]{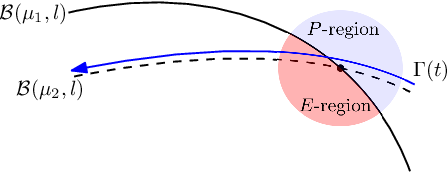}
\caption{$\mathcal{B}_{\mu_1,l}\cap\mathcal{B}_{\mu_2,l}\neq \O$ leads to a trajectory $\Gamma(t)$ that violates semipermiability} \label{semiPerm}
\end{figure}

\emph{Proof of Theorem~\ref{thm:barrier}:} Let us assume that $\mathcal{B}(\mu_1,l)$ intersects $\mathcal{B}(\mu_2,l)$ at some $(x_I, y_I)\in \mathds{R}^2$. From Lemma \ref{semipermeableBarrier}, $\mathcal{B}(\mu_1,l)$ is a semipermeable surface.  Hence, at any point above the curve $\mathcal{B}(\mu_1,l)$, if $P$ plays optimally, it is not possible for $E$ to drive the state across the barrier. If $\mathcal{B}(\mu_2,l)$ intersects  $\mathcal{B}(\mu_1,l)$, then in the vicinity of the intersection point, there exists a strategy of $E$ that can drive the state across $\mathcal{B}(\mu_1,l)$ while $P$ plays its optimal strategy. An example of such a trajectory (say $\Gamma(t)$) is shown in Fig. \ref{semiPerm}. This violates the semipermeability property of $\mathcal{B}(\mu_1,l)$  and contradicts our initial assumption that $\mathcal{B}(\mu_1,l)$ intersects $\mathcal{B}(\mu_2,l)$. This completes the proof of Theorem~\ref{thm:barrier}. \qed

\section{Asymmetric Information Homicidal Chauffeur Game with Deceptive Evader}
We will now consider an asymmetric information version of HCG, where the following assumptions hold.

\medskip
\begin{assumption}
    The evader has two discrete choices of speeds, namely $\mu_1$ and $\mu_2$, where $\mu_1>\mu_2$.
\end{assumption}
\begin{assumption}
     $P$ does not know the bound on the evader's speed, $\bar{\mu}$, but can observe it as $\hat{\bar\mu}(t)$ based on evader's action during the game using the following measurement model,
\begin{align}\label{measurementModel}
    \hat{\bar\mu}(t) &= \underset{\bar{t} \in [0,t]}{\sup} ~u(\bar{t}), &\forall t\in(0,t_f).
\end{align}\end{assumption}

\medskip

That is, at any instance of time, $P$ estimates a bound on the speed as the maximum speed at which the evader has moved so far in the game.
%

\medskip

\begin{assumption}
$E$ has complete information about the game states and the parameters, and is aware of the information disadvantage of $P$.  
\end{assumption}
\begin{assumption} \label{SwitchOnceAssump}
    $E$ can only switch once from a lower to a higher speed. 
\end{assumption}

\medskip

Under these assumptions, the game reduces to an asymmetric-information variant of the HCGVSE, with the evader’s speed restricted to discrete choices $\mu \in {\mu_1,\mu_2}$ rather than a continuous range $\mu \in [0,\bar{\mu}]$. This simplification both facilitates analysis and highlights a clear instance of deception, while the continuous-speed version will be addressed in future work. Assumption \ref{SwitchOnceAssump} is also adopted for tractability; allowing switches from higher to lower speed would require computing the equivocal surface for the HCGVSE, which may differ from the standard HCG since maintaining equivocation could involve reducing speed—an extension we defer to later work.

\medskip

\begin{figure}[!htbp]
    \centering
    \includegraphics[width=0.4\textwidth]{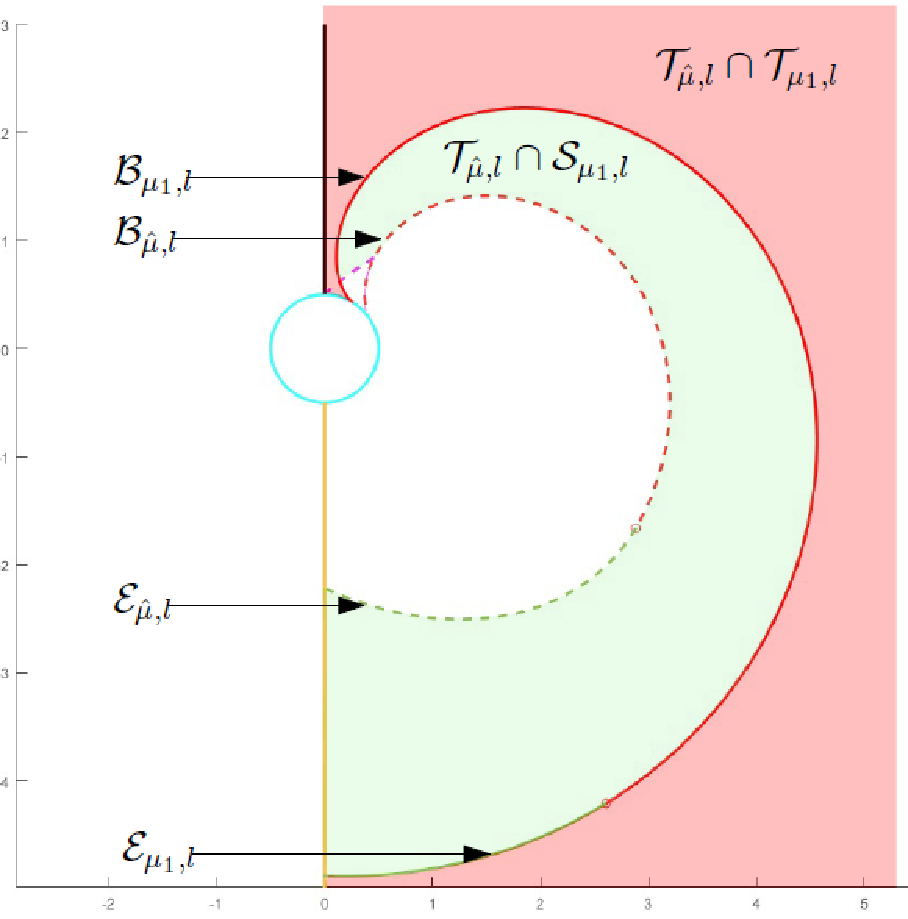}
    \caption{Superposition of HCG with two speed parameters $(\hat{\mu},l)$ and $(\mu_1,l)$}
    \label{fig:shaded}
\end{figure}

The game may thus be viewed as a superposition of two HCGs with speed parameters $\mu_1$ and $\mu_2$. While the pursuer acts according to $(\hat{\mu},l)$, the evader, aware of the pursuer’s disadvantage, exploits both $(\hat{\mu},l)$ and $(\mu_2,l)$. By Theorem \ref{thm:barrier}, two regions of interest emerge from this superposition: These need to be subscripts $\mathcal{T}_{\hat{\mu},l}\cap \mathcal{T}_{\mu_1,l}$ and $\mathcal{T}_{\hat{\mu},l}\cap \mathcal{S}_{\mu_1,l}$, illustrated in Fig.~\ref{fig:shaded}. The results below analyze the potential for deception within these regions.

\begin{remark}\label{noConser1}
If $\hat{\bar{\mu}} < \bar{\mu} $ and $\left(x_0,y_0\right)\in \mathcal{T}_{\hat{\bar\mu},l}\cap \mathcal{S}_{\bar\mu,l}$,  then the pursuer will deploy a sub-optimal control and hence a higher capture time compared to the case when $\hat{\bar{\mu}} = \bar{\mu} $ will be achieved. 
\end{remark}

\medskip

\begin{remark}\label{noConser2}
If $\hat{\bar{\mu}} < \bar{\mu} $ and $\left(x_0,y_0\right)\in \mathcal{S}_{\hat{\bar\mu},l}\cap \mathcal{T}_{\bar\mu,l}$,  then the pursuer will again deploy a sub-optimal control and hence a higher capture time compared to the case when $\hat{\bar{\mu}} = \bar{\mu} $ will be achieved. 
\end{remark}

\medskip

Remarks \ref{noConser1} and \ref{noConser2} emphasize that the precise knowledge of the bound on the evader's speed is necessary for the pursuer to capture the evader at the optimal time. Having a conservative estimate $\hat{\bar{\mu}} = 1-\epsilon$, where $\epsilon$ is infinitesimally small does not help the pursuer.

We now introduce a class of evader strategies that aims at exploiting the information disadvantage of $P$. 

\medskip

\begin{definition}[Deceptive speed strategy $\mu_d(t)$] A strategy is called a \emph{deceptive speed strategy} if $E$ chooses a speed 
    $$\mu_d(t) = \begin{cases}
    \mu'(t)< \mu^*(t), & t\in[0,\bar{t}], \\
     \mu^*(t), & t\in[\bar{t},t_f], \\
    \end{cases}$$ for some $\bar{t}<t_f$ with the intent to bias $P$'s estimate $\hat{\mu}(t)=\mu'(t)$, $\forall t\in[0,\bar{t}]$.
\end{definition}

For the evader, a deceptive speed strategy is called an \textit{advantageous deceptive strategy} if 
\begin{align}
J\left(u^*,\psi^*,\mu^*\right)< J\left(u^*,\psi^*,\mu_d\right).
\end{align}

The following result characterizes the region in the game space of the HCGVSE for which there does not exist an advantageous deceptive strategy for the evader.

\medskip
\begin{figure*}[htbp]
    \centering
    \subfigure[]{\includegraphics[width=0.3\textwidth]{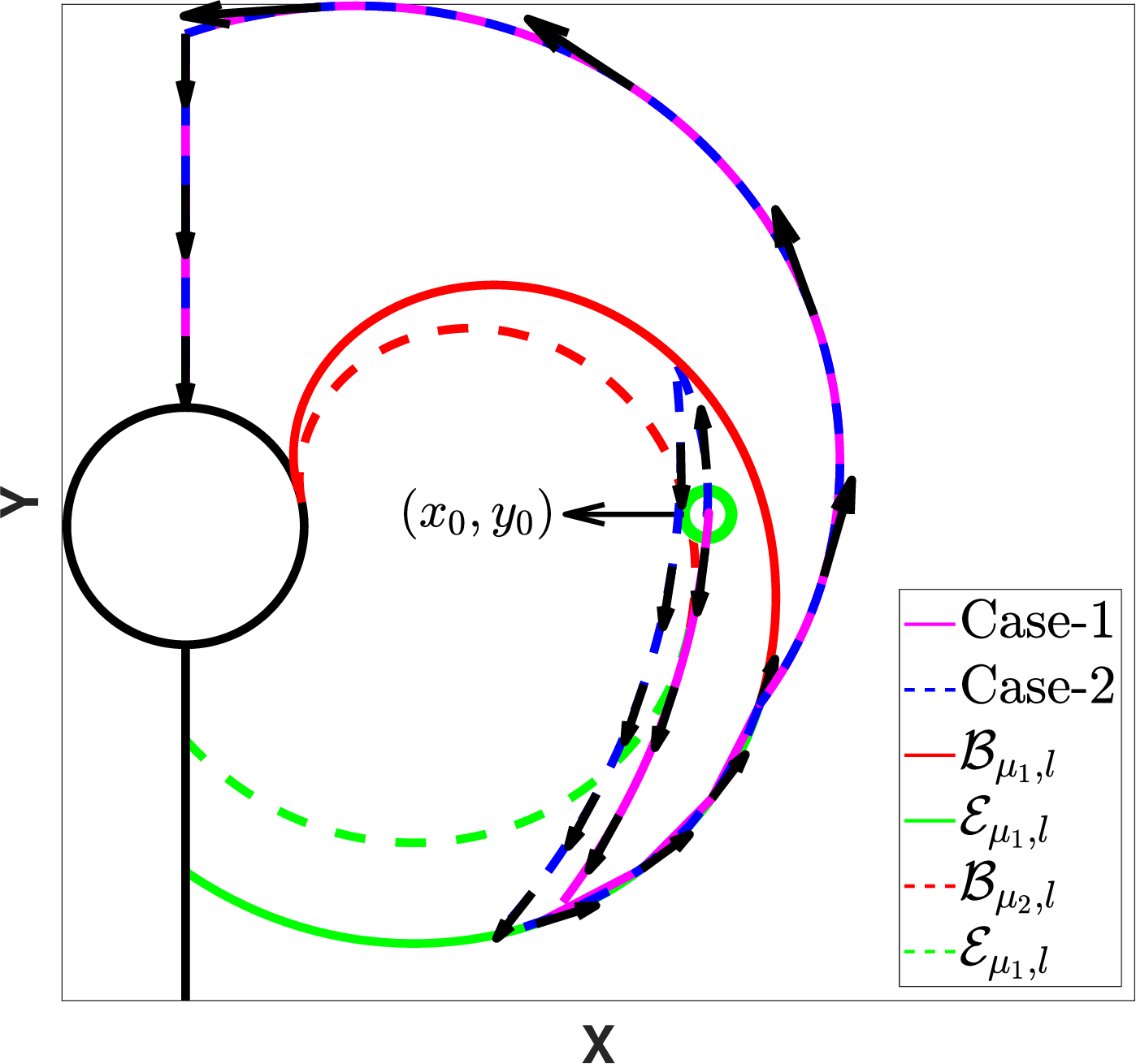}\label{figdec:a}}
    \subfigure[]{\includegraphics[width=0.3\textwidth]{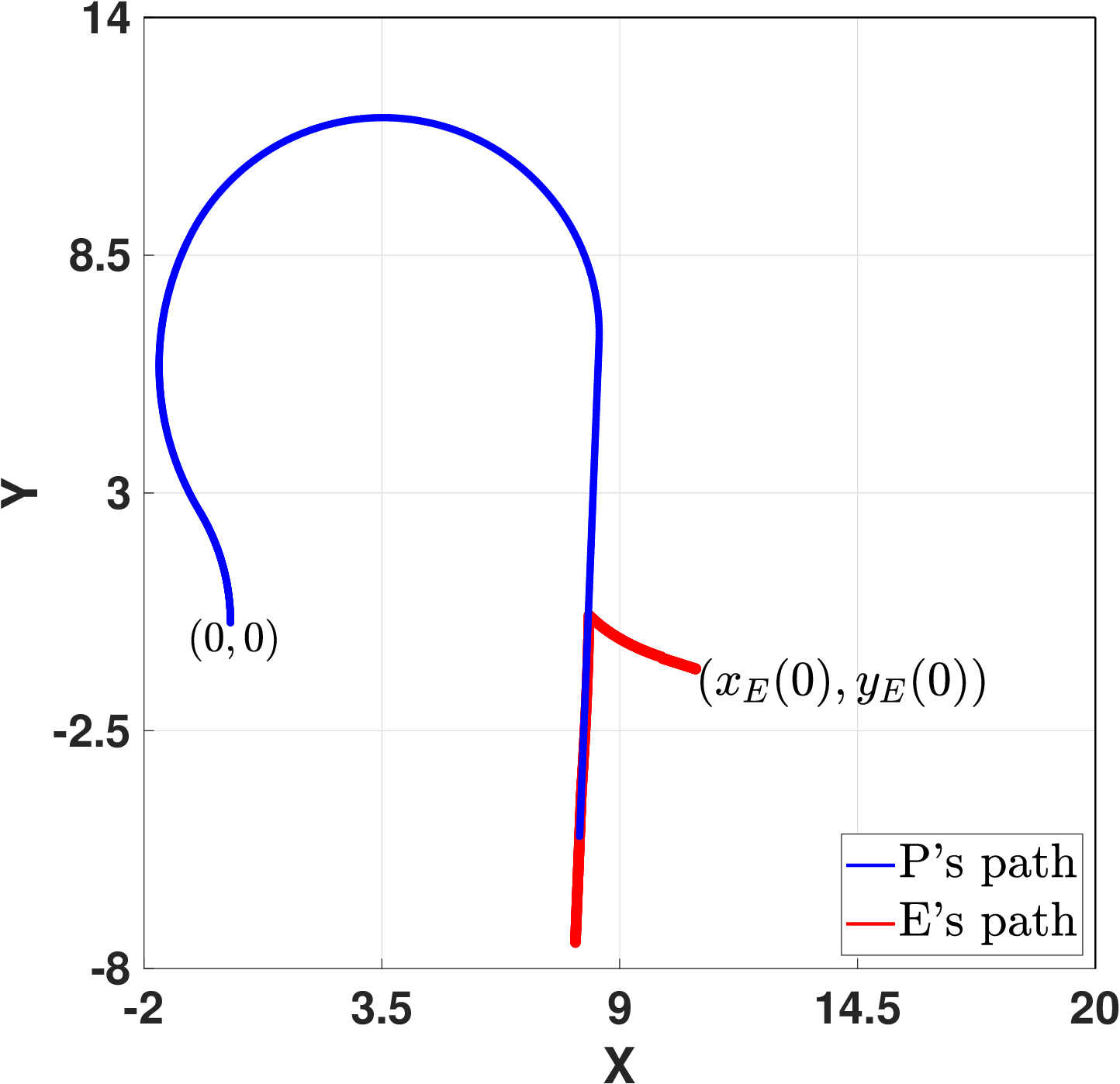}\label{figdec:b}}
    \subfigure[]{\includegraphics[width=0.3\textwidth]{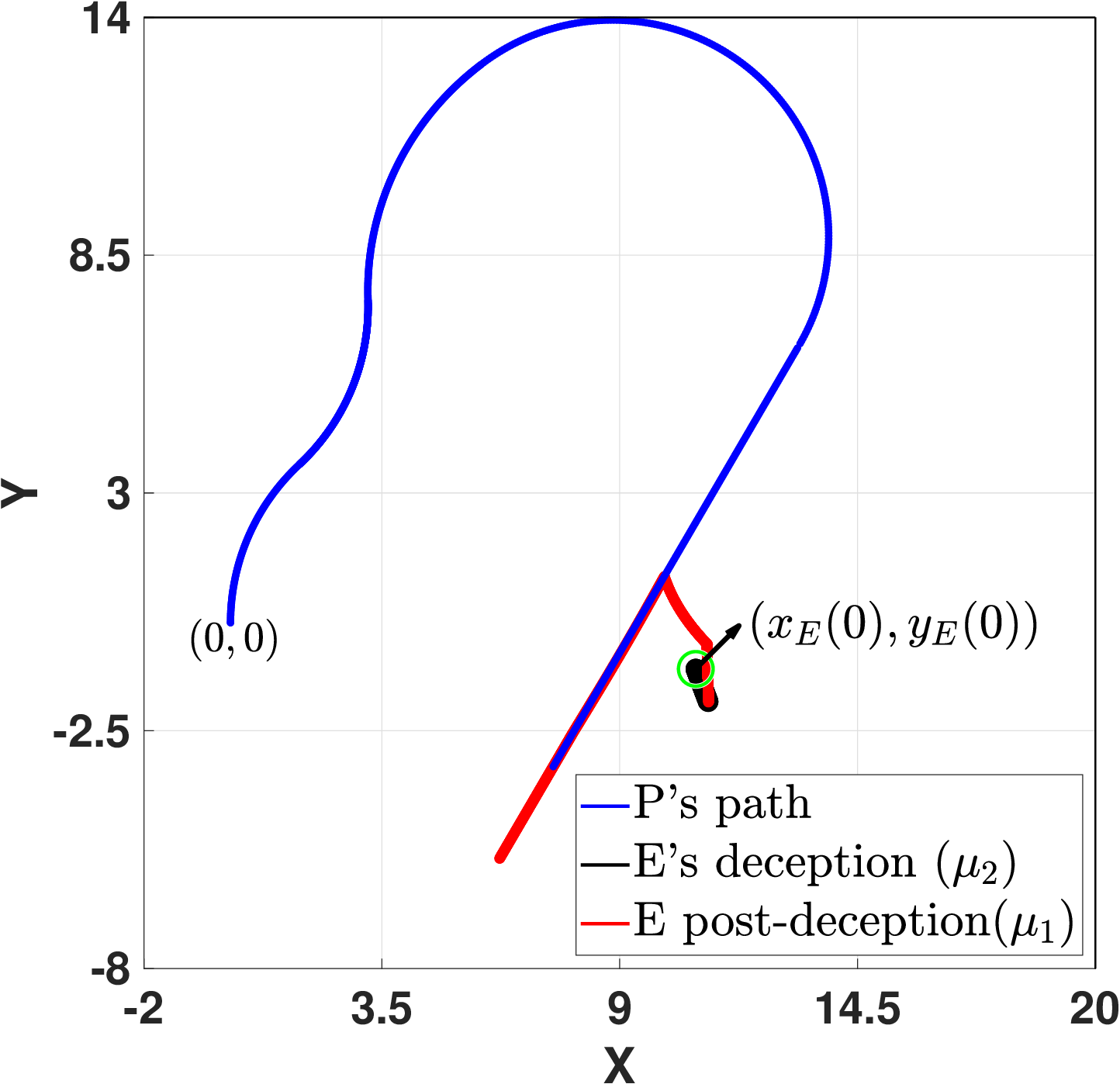}\label{figdec:c}}
    \caption{\textbf{a)} Deception in HCG shown in the pursuer-relative coordinate system. \textbf{b)} Case 1: $E$ (red) moves with $\mu_1 = 0.3$ and $P$ (blue) responds optimally. \textbf{c)} Case 2: $E$ initially moves with $\mu_2 = 0.2$ (black) to deceive $P$, and switches to $\mu_1 = 0.3$ (red) upon reaching the barrier corresponding to $\mu_1$.}
\end{figure*}

\begin{theorem}\label{thm:nodeception} If $(x_0,y_0)\in \mathcal{T}_{\mu_1,l}\cap \mathcal{T}_{\mu_2,l}$ for any $\mu_1>\mu_2$, then there does not exist an advantageous deceptive strategy for $E$.
\end{theorem}
\noindent Proof: The proof is based on applying Lemma \ref{csdub} and Theorem~\ref{thm:barrier}. Suppose that in the duration $[t, t+\Delta t]\cap[0,t_{UL}]$, $E$ travels with a speed $\mu_2<\mu_1$ to deceive $P$. But from Lemma~\ref{csdub}, we conclude that $P$ would still take $t_{UL}$ to head along $E$. Now, the separation between $P$ and $E$ is given by
\begin{align*}
    s = s_0+\mu_1 \cdot t +\mu_2 \cdot \Delta t+\mu_1 \cdot (t_{\text{UL}}-(t+\Delta t))\\
    = s_0+(\mu_2-\mu_1) \cdot \Delta t+\mu_1 \cdot t_{\text{UL}}.
\end{align*}  Since $(\mu_2-\mu_1)<0$, $E$ ends up reducing the separation and as a result, also ends up with a loss in time on the $\mathcal{U}^+_{\mu,l}$. This makes the attempt to deceive $P$ result in a suboptimal cost for the evader, and thus, there does not exist an advantageous deceptive strategy for $E$. \qed

\medskip

The next result shows evidence of a region in the game space from which there exists an advantageous deceptive strategy for the evader.

\medskip
\begin{proposition}\label{thm:deceptive}
For some $(x_0,y_0)\in \mathcal{S}_{\mu_1,l}\cap \mathcal{T}_{\mu_2,l}$ and $\mu_1 > \mu_2$, there exists an advantageous deceptive strategy for the evader $E$.
\end{proposition}

\noindent\textit{Proof:} We construct a numerical example that provides evidence of an advantageous deceptive strategy for $E$ when $(x_0,y_0)\in \mathcal{S}_{\mu_1,l} \cap \mathcal{T}_{\mu_2,l}$. As a representative scenario, consider the following setup:

Let the evader $E$ have two possible speed choices: $\mu_1 = 0.3$ and $\mu_2 = 0.2$. Set the capture radius $l = 0.5$. The initial positions are: $(x_{P0}, y_{P0}) = (0, 0), \quad (x_{E0}, y_{E0}) = (10.76, -1.07),
$ which corresponds to a pursuer-relative initial condition of 
$
(x_0, y_0) = (2.152, -0.214) \in \mathcal{S}_{\mu_1,l} \cap \mathcal{T}_{\mu_2,l},
$
as shown in Fig.~\ref{figdec:a}.

\begin{enumerate}
\item \textbf{Case 1 – Baseline (Full Information) Strategy:}
With full knowledge of the evader’s maximum speed $\mu_1$ with $P$, $E$ follows the equilibrium HCG strategy for $(\mu_1,l)$: it proceeds along a secondary trajectory in $\mathcal{S}_{\mu_1,l}$, followed by the equivocal curve $\mathcal{E}_{\mu_1,l}$, then a tributary trajectory in $\mathcal{T}_{\mu_1,l}$, and finally along the positive universal line $\mathcal{U}^+_{\mu_1,l}$ until capture. Capture occurs in 17 s (Fig.~\ref{figdec:a} (magenta solid line), and the trajectory in the global $X-O-Y$ coordinates is illustrated in Fig.~\ref{figdec:b}.).

\item \textbf{Case 2 – Deceptive (Asymmetric Information) Strategy:}  
When $P$ is unaware of $\mu_1$, $E$ initially moves at $\mu_2=0.2$, selecting $\psi^*(t)$ to mimic the $(\mu_2,l)$ tributary trajectory. Observing this, $P$ commits to the corresponding strategy. Once the trajectory in the relative frame reaches the barrier $\mathcal{B}_{\mu_1,l}$ at some point $(x_b, y_b)$, $E$ is compelled to switch to the true maximum speed $\mu_1$ and follows the $(\mu_1,l)$ equilibrium path (secondary–equivocal–tributary–universal). Capture occurs in 20 s (blue dashed line in Fig.~\ref{figdec:a} and Fig.~\ref{figdec:c}).

\item \textbf{Result:}
The deceptive strategy in Case 2 delays capture by about 3 s relative to Case 1. This confirms that $E$ can benefit from deception when starting in $\mathcal{S}_{\mu_1,l} \cap \mathcal{T}_{\mu_2,l}$, completing the proof. 
\end{enumerate}

This result suggests the possibility of existence of numerous similar deception strategies, which will be investigated further in future research.

\section{Conclusion and Future Directions} This paper analyzed a novel variant of the classic Homicidal Chauffeur differential game under asymmetric information, where the pursuer is unaware of the evader’s maximum speed. Instead, the pursuer estimates this bound in real time based on the evader’s observed motion. We showed that this asymmetry enables deceptive strategies in which the evader initially moves slowly to mislead the pursuer, then switches to its true speed to prolong capture. In establishing these strategies, we derived structural properties of an HCG with variable evader speeds. Our analysis further showed that deception is advantageous only when the evader starts in the region bounded by the barriers of two HCGs with different speeds; in regions governed by tributary trajectories, deception provides no benefit. {A precise characterization of the regions in the game space that admit deceptive strategies will be a part of future work.}

\bibliographystyle{ieeetr}
\bibliography{main}

\end{document}